# Flux noise in ion-implanted nanoSQUIDs

Giuseppe C. Tettamanzi,[1,2*] Christopher I. Pakes[3], Simon K. H. Lam,[4], Steven Prawer[1]

[1]*School of Physics, University of Melbourne, Victoria 3010, Australia.*

[2] *Kavli Institute of Nanoscience, TU Delft, Lorentzweg 1, 2628 CJ DELFT, The Netherlands.*

[3]*Department of Physics, La Trobe University, Victoria 3086, Australia.*

[4]*CSIRO, Materials Science and Engineering, PO Box 218, Lindfield, NSW 2070, Australia.*

\* Corresponding author; Email: G.C.Tettamanzi@tudelft.nl

Abstract:

Focused ion beam (FIB) technology has been used to fabricate miniature Nb DC SQUIDs which incorporate resistively-shunted microbridge junctions and a central loop with a hole diameter ranging from 1058 nm to 50 nm. The smallest device, with a 50 nm hole diameter, has a white flux noise level of $2.6\ \mu\Phi_0/\text{Hz}^{1/2}$ at $10^4$ Hz. The scaling of the flux noise properties and focusing effect of the SQUID with the hole size were examined. The observed low-frequency flux noise of different devices were compared with the contribution due to the spin fluctuation of defects during FIB processing and the thermally activated flux hopping in the SQUID washer.

## 1. Introduction

DC Superconducting Quantum Interference Devices (SQUIDs), operating as a flux-to-voltage transformer, have found many applications in sensitive magnetic field detection [1]. Over the past 40 years, SQUIDs that have been used in different applications implemented Josephson junctions with barriers of a range of different materials [2], as well as based on microbridge junctions [3, 4]. Recently, there is a strong interest for using SQUIDs in the application of detecting small spin systems in nanometer scale using microbridge junctions [5]. Further miniaturisation of the devices could potentially lead to measurement of the magnetic properties of systems demonstrating single electron spin [5-7], where a variation in the magnetic moment of a sample situated within the SQUID loop was measured in the presence of the device's own flux noise. This may be realised by the development of nanotechnology instrumentation that can reproducibly make structures in submicron sizes. Formation of a Josephson junction using a nano-sized microbridge is expected to possess low intrinsic capacitance appropriate for low-noise applications as there is no barrier structure in the junction [8]. A figure-of-merit has been identified [5] to achieve sufficient spin sensitivity for single spin measurement. This figure-of-merit is linearly scaled with the size of the SQUID hole and the flux noise of the device [5]. Approaches for the realisation of small microbridge SQUIDs using electron beam [5,9] and atomic force [10] lithography have limitations for the achievable minimum feature size and the reproducibility, limiting the initial promise of these technologies. An alternative approach of employing focused ion beam (FIB) technology to engineer Nb SQUIDs by ion-beam milling has recently gained popularity [11, 12], and devices with a physical hole diameter to approximately 200 nm have been demonstrated [11]. Reports in the literature, describing the fabrication of microbridge-based SQUIDs with sub-micron hole diameter, have focused on understanding the level of white flux noise present, which is dominated by the Johnson noise of the shunting resistor of the microbridge junction. The shunting element could be an external shunting resistor [9, 12-14] or the damaged superconducting material during the fabrication process [11]. DC SQUIDs made by FIB techniques have been reported by Troeman [11]. The white noise level is approximately $1.5\,\mu\Phi_0/\mathrm{Hz}^{1/2}$, giving a spin sensitivity of a few tens of electron spin, with a $1/f$ noise component extending to few hundreds Hertz. Further improvement on white noise was recently demonstrated by Hao [12] using a second stage SQUID amplifier and a W protective layer on the SQUID. They achieved a white noise level of $0.2\,\mu\Phi_0/\mathrm{Hz}^{1/2}$, again with a $1/f$ noise component extending to approximately 1 kHz.

We are particularly interested in exploring the regime of the hole size below 100 nm, for which a single-spin flip in a magnetic sample within the SQUID hole may be detectable. To achieve the goal of measurement of a single molecular spin, it is important to elucidate the origin of $1/f$ noise in DC SQUIDs made by FIB technique. Here, we report the noise properties and effective area of our FIB DC SQUIDs with different SQUID hole diameter, aiming at getting better insight of the noise properties of FIB fabricated devices and the focusing effect of the SQUID washer. A important recent theoretical report from Koch *et al.* [15] has examined the $1/f$ noise contribution due to electron trapping and hopping on defect sites within the underlying substrate. FIB engineering of SQUIDs introduces a significant number of defects in the SQUID hole and the edge of the Nb structure to a lateral dimension of about 30 nm [11]. By examining the scaling of the noise properties with SQUID hole size, the impact of deliberately introduced defects in contributing to $1/f$ noise in FIB DC SQUIDs is considered and compared to noise due to vortex hopping in the Nb washer.

## 2. Experimental Details and Results

The devices were fabricated on Nb films with 30 nm thickness, grown on $SiO_2$/Si substrates in a DC sputter magnetron system. An in-situ deposited 30 nm Au film on top of the Nb layer provided a high shunt resistance across the fabricated junction in the final device [5]. Patterning of the films was carried out by FIB milling using 30 keV Ga-ions in a crossed-beam FIB/SEM (Scanning Electron Microscope) system. The characterisation of five Nb SQUIDs is presented in Table 1. All SQUIDs had a square washer with an area of 1500 μm$^2$. The SEM provides imaging of the fabricated device to monitor the milling process and to determine the SQUID loop and washer dimensions (Figure 1). Utilising a simple fabrication protocol, the FIB was also used to pattern Au/Nb bonding pads for electrical connection to the device. The physical width of each nanobridge is typically about 200 nm, but due to the effect of Ga-ion poisoning in its periphery, an active superconducting filament of about 120 nm in width exists within the centre of the nanobridge [11].

Transport measurements of bonded devices were performed over a range of temperatures down to 4.2 K using Star Cryoelectronics DC SQUID drive electronics with nV/√Hz sensitivity [16]. Current-voltage (*I-V*) curves were obtained for all devices as a function of temperature. Figure 2a illustrates typical *I-V* data, obtained for the smallest device (A) at a temperature of 6 K. At temperatures of 6 K and above, the *I-V* characteristics of all devices were found to be non-hysteretic. At lower temperatures, the critical current of the junctions was of the order 1 mA, and likely to give rise to significant heating of the junction in its normal state leading to hysteretic behaviour. To determine the flux-to-voltage transfer function $(\partial V/\partial \Phi_0)$ of the devices, the SQUID response to an AC magnetic field of about 8 Hz, applied using an external field coil, was studied. Figure 2b shows the response to the applied field of device A, obtained at different bias currents. The optimum bias current was found by maximising the output voltage. The maximum output voltage and the transfer function were measured as a function of the applied magnetic field. In order to calculate the magnetic field to magnetic flux (coupled into the SQUIDs) conversion, the periodicity of the voltage – applied magnetic field modulation (*V-B*) was measured for each device. The effective area of the devices, $A_{\text{eff}}$, is defined as the value of magnetic field to give one flux quantum ($\Phi_0$) modulation on the *V-B* characteristic and is shown in Table 1. For our smallest device (A), the maximum applicable field in the measurement apparatus corresponds to an applied field that is considerably smaller than $\Phi_0$ through the SQUID loop. For this device, the transfer function has been estimated, based on the modulation behaviour of the devices with larger hole dimension. The measured value of $A_{\text{eff}}$ for all the devices was approximately two orders of magnitude larger than the area of the SQUID hole, indicating the device's washer has flux focussing. The SQUID washer focusing effect was studied 20 years ago by Ketchen *et al.* [17]; $A_{\text{eff}}$ is expected to scale linearly with the hole dimension of the device, which is in contrast to our data given in Figure 3. However, the calculation presented by Ketchen *et al* assumed that the penetration depth of the superconducting material is less than the half of the film thickness, *t*. This assumption is not applicable for the nanoSQUIDs discussed here as the film thickness of $t = 30$ nm is much less than the penetration depth, $\lambda \sim 100$ nm, of Nb [18]. Rather, the effective penetration of thin film, $\lambda_{\text{eff}} = 2\lambda^2/t$, is about 1 $\mu$m, indicating that the screening current in the washer has a much larger spatial distribution than a thick film. Further detailed modelling of the flux focusing effect of very thin film devices is required to fully explain these experimental results.

The spectral response of the voltage noise, $V_n$, was measured for all devices using a spectrum analyser at temperatures in the range of 4.33 to 6 K for which each device exhibited non-hysteretic behaviour. The corresponding flux noise, $\Phi_n = V_n/(\partial V/\partial \Phi_0)$, was calculated in each case. Table 1 summarises the transfer function and the flux noise, $\Phi_n$, obtained for all devices at 1 Hz. For illustration, the spectral flux noise density obtained for devices A, C and E are shown in Figure 4. Peaks in the noise spectra at low frequency originate from the AC magnetic field and its harmonic components. The smallest device (A) was found to exhibit a flat noise response above about 100 Hz, with a voltage noise level close to the expected noise floor ($\text{nV}/\text{Hz}^{1/2}$) of the measurement amplifier. It is instructive to consider the flux noise as a useful figure-of-merit for magnetic measurement applications. The smallest device (A), with 50 nm hole size, has a white noise value of $\Phi_n = 2.6\,\mu\Phi_0/\text{Hz}^{1/2}$. This is comparable to devices reported in the literature, with flux noise figures in the range of $1.5 - 7.0\,\mu\Phi_0/\text{Hz}^{1/2}$ [5, 9, 10]. The flux noise in Device A is amplifier limited, and improved performance is anticipated with the use of a low-temperature SQUID amplifier. This arrangement has recently demonstrated a magnetic flux sensitivity of 0.2 $\mu\Phi_0/\text{Hz}^{1/2}$ for a device with loop dimension of 200 nm fabricated via optical and FIB processing, and measured using a SQUID amplifier array [12]. Importantly, the simplified FIB-based fabrication protocol described in this paper is well-suited to extend this approach for the development of a 50 nm nanoSQUID sensor with an *on-chip*, conventional SQUID amplifier, where the amplifier has been fabricated in a multi-layer architecture on the substrate prior to FIB patterning of a nanoSQUID at the amplifier input.

All the devices described here exhibited an excess noise with white noise being dominant (1/f noise corner) at different frequencies. The noise level at 1 Hz is shown in Table 1. The $1/f$ corner frequency of device A is about 100 Hz while other devices show a $1/f$ noise corner exceeding $10^4$ Hz. As the bandwidth of the existing experimental measurement set up has a cut-off frequency of ~ 30 kHz, the white noise of devices B-E were not exactly observed. Below is a discussion of what may be contributing to this excess flux noise.

## 3. Estimation of contribution to the flux noise

### a. Low frequency noise due to electron hopping

The noise contribution due to a magnetisation change of a population of randomly positioned defects introduced within the SQUID hole can be estimated using a simple analytical model based on the calculation of the net-flux threading the SQUID loop. It is instructive to first consider the net-flux threading the loop due to an isolated spin of moment $\mu_B$ situated at the centre of the SQUID loop, as considered by Ketchen [6]. A change in the magnetic moment, $\mu_B$, arising from a change in spin orientation, gives rise to a net-flux threading the SQUID loop with a hole radius, $a$, which takes the form $\Delta\phi = \int_0^{2\pi}\int_a^{\infty} \left(\mu_0 \mu_B / 4\pi r^3\right) r dr d\theta$, where $r$ is the radial distance from the centre of the loop. Considering a spin positioned within the hole, but at a distance $y$ from the centre of the device, the magnetic flux threading the SQUID loop is of the form

$$\Delta\phi = \frac{\mu_0}{4\pi} \int_0^{2\pi}\int_a^{d} \mu_B \left(r^2\sin^2\theta + (r\cos\theta - y)^2\right)^{-\frac{3}{2}} r dr d\theta, \qquad (1)$$

where $r$ is the radial distance from the centre of the loop and $\theta$ is the angle between the **r** and **y** directions. The limit, $d$, is taken to be $100\mu m$, extending significantly beyond the SQUID washer. It follows that the flux noise [15] for a large number of defect sites, distributed randomly throughout the SQUID hole with areal density $n$, will be of the form

$$\Phi_n(f) = \left( \int_0^{a+s} \left[ \frac{\mu_0}{4\pi} \int_0^{2\pi} \int_{a+s}^d \mu_B \left( r^2\sin^2\theta + (r\cos\theta - y)^2 \right)^{-\frac{3}{2}} r \, dr \, d\theta \right]^2 n 2\pi y \, dy \Big/ 30 f \right)^{\frac{1}{2}}. \qquad (2)$$

To account for the introduction of defects into the periphery of the SQUID hole, the radius of the Ga-ion implanted region is taken to be $a + s$, where $s$ is the lateral ion straggling distance. The areal defect density is considered to be constant within this region, so an approximate reduced value of $s = 20$ nm is used. We shall examine the dependence of the flux noise given in equation (2) on the SQUID hole diameter, using the areal defect density as a variable parameter at a frequency of 1 Hz. For a device with a hole radius 50 nm, inclusion of the term $s$ in equation (2) was found to approximately double the value of the calculated flux noise. However, for larger devices, defects in the periphery of the SQUID hole do not contribute a significant noise level beyond that due to the total number of defects introduced into the hole.

SRIM simulations [19] were used to estimate the areal density of defects in the SQUID hole following Ga-ion milling of the Nb/Si substrate. Ga ions sputter the Nb material in the centre of the SQUID loop with a sputter yield of about 5 for Nb, which has a number density of $N = 1.4 \times 10^{26}$ atoms.m$^{-3}$. In addition, for each incoming Ga ion implanted with 30 keV energy, approximately 650 vacancies are created at a depth $z \sim 20$ nm from the ion entry site on the surface of the substrate. On completion of the milling process, the bottom of the SQUID hole will contain defects to a depth of about 20 nm in the underlying substrate. The areal density of defects will therefore be of the order of $n \sim 650 z N/5 = 3 \times 10^{20}$ m$^{-2}$. The variation in spectral flux noise response with SQUID hole diameter given in equation (2), calculated for a SQUID hole occupied by spins of moment $\mu_B$ with this areal density, is illustrated by the solid line in Figure 5. A contribution to the device flux noise will also arise from randomly orientated defects in the unpatterned regions of the underlying substrate [15]; which are expected to have an areal density of about $5 \times 10^{17}$ m$^{-2}$ [20]. The broken line shown in Figure 5 illustrates the calculated flux noise coupled into the SQUID hole from spins of moment $\mu_B$ in the SQUID washer at this areal density. It is clear from the figure that the contribution to the measured flux noise from defects within the washer is insignificant for our devices, compared to those examined in the literature [20], due to the significantly smaller loop hole diameters achieved by FIB processing.

The calculated spectral noise response due to defects within the SQUID hole is consistent (within a factor of two) with the measured noise level for devices A and B. Given uncertainties in estimation, using SRIM, of the number of defects introduced into the device by Ga-ion implantation, and the extent to which defects are introduced into the periphery of the SQUID hole, electron hopping seems to be a plausible model to explain the noise in these devices. However, this model fails to account for the observed dependence of the measured flux noise on the SQUID hole size, and we seek an alternative model to explain the rapid increase in spectral noise response as the SQUID hole size extends beyond 200 nm.

**b. Low frequency noise due to hopping vortices**

Although the devices were cooled in a low magnetic field environment, magnetic vortices can still penetrate into a superconducting thin film with poor quality edges where vortices tend to nucleate. In our experiment, the residual field was in the order of 0.5 $\mu$T as a consequence of a single Mu-metal magnetic shield. The number of vortices, $N$, that penetrate into the washer would depend on the stray field that the SQUID experienced during the cool down through its transition temperature, which is expected to be proportional to the value of the stray field. In our case, the stray field was the same for all the devices as they are on the same chip during the measurements. However, the flux focussing effect of the SQUID washer (given by the effective area) would result in an average magnetic induction of the order of 20 $\mu$T (e.g., device A), experienced in the SQUID hole. The edges of the hole would experience even larger values. Therefore, the actual field intensity on the material near the SQUID-hole edges would be "amplified" by the focusing effect, which is governed by the effective area, $A_{\text{eff}}$, of each device, and $N$ is expected to be linearly proportional to $A_{\text{eff}}$. The dependence of low frequency noise against the number of trapped vortices in a YBCO SQUID has been studied by Doenitz et al. [21]. The spectral noise response at 1 Hz, $\Phi_n^2$(1 Hz) in the unit of $\Phi_o^2$/Hz, was found to have a dependence of $N^{1.5}$. In our experiments, the number of penetrated vortices is unknown. However, a qualitative analysis can be performed using the measured value of $A_{\text{eff}}$. As discussed above, we expect $N$ to be proportional to $A_{\text{eff}}$, and therefore $\Phi_n^2$(1 Hz) would be proportional to $A_{\text{eff}}^{1.5}$. Figure 6 shows a plot of $\Phi_n^2$(1 Hz) against $A_{\text{eff}}$ of our devices listed in Table 1. Although there is scattering of the data, the fit is reasonable good to a power law of 1.5.

These results suggest there are large numbers of vortices being trapped near the SQUID hole even at very low stray fields. Experiments have shown that improving the $J_c$ value of the SQUID edges greatly reduces the low frequency noise [22, 23]. As discussed above, the edges of the Nb structure in FIB-engineered devices would incorporate Ga-ion implanted damage in their periphery. These regions would become non-superconducting, or have lower $T_c$ and $J_c$ values compared with the unexposed area, depending on the Ga-ion concentration. Therefore, the device is likely to have a layer of low-$J_c$ material along the edges that would be subject to the nucleation of vortices even at low magnetic fields. This suggests that ion-beam induced damage arising from FIB processing of nanoSQUIDs contributes to their characteristic $1/f$ flux noise, predominantly through modification of the superconducting properties of Nb material in the periphery of the Nb structure. Further improvement of the FIB process would be needed to address this problem, including the use of a more robust layer of material for a resistive shunt to minimise ion-beam damage to the underlying Nb structure [12], and post-FIB thermal annealing.

**4. Conclusion**

In summary, focussed ion beam processing has been used to fabricate Nb SQUID devices, incorporating microbridge junctions, with SQUID loop hole sizes in the 50 to 1000 nm range. The scaling of low frequency, $1/f$ flux noise with device hole dimension has been measured and compared with the expected noise contributions due to electron hopping on defects introduced by FIB-processing, and hopping of vortices within the SQUID washer. Our analysis suggests that $1/f$ noise predominantly arises in FIB-engineered devices due hopping of vortices within the SQUID washer, with an enhancement in vortex activity likely in the ion-beam damaged region at the periphery of the SQUID hole.

This work was supported by the Australian Research Council under the Centres of Excellence scheme. G.C. Tettamanzi acknowledges the Dutch Foundation for Fundamental Research on Matter (FOM) for financial support. We thank A. Potenza for providing sputtered Nb films, C. Granata for useful discussion, and C. Foley for commenting on the manuscript.

| Device | A | B | C | D | E |
|---|---|---|---|---|---|
| Physical Hole Diameter (nm) | 50 | 168 | 258 | 600 | 1058 |
| Effective Area, $A_{eff}$ ($10^{-12} m^2$) | 0.08 | 0.828 | 2.07 | 17.25 | 103.5 |
| Transfer Function, $\partial V/\partial \Phi_0$ ($\mu V/\Phi_0$) | 600 | 1000 | 100 | 400 | 300 |
| Voltage Noise, $V_n$ at $10^4$ Hz (nV/Hz$^{1/2}$) | 1.5 | 11.68 | 1.6 | 13.69 | 19.54 |
| Flux noise, $\Phi_n$ at 1 Hz ($\mu\Phi_o$/Hz$^{1/2}$) | 5.0 | 35 | 110 | 700 | 1020 |

**Table 1:** Summary of SQUID characteristics reported in this study.

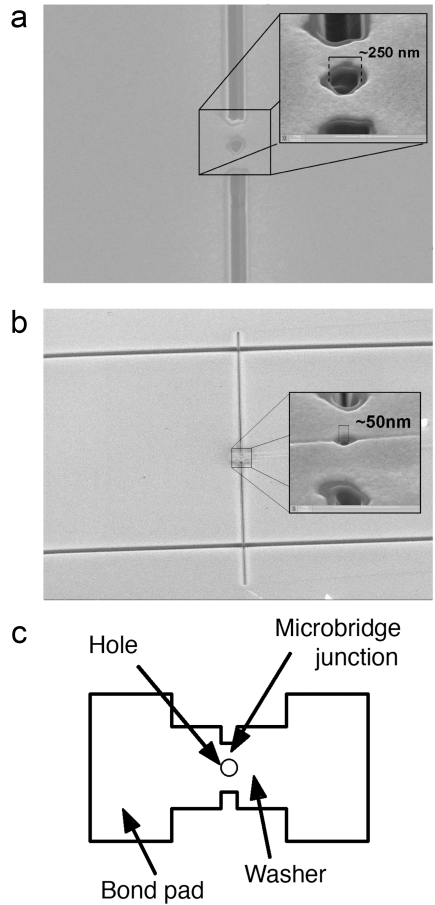

**Figure 1:** Scanning electron micrograph of (a) device C and (b) device A. In each case, the image is enlarged in the inset to show in detail the nanobridge junctions and hole in the SQUID loop. (c) Schematic illustration of device structure.

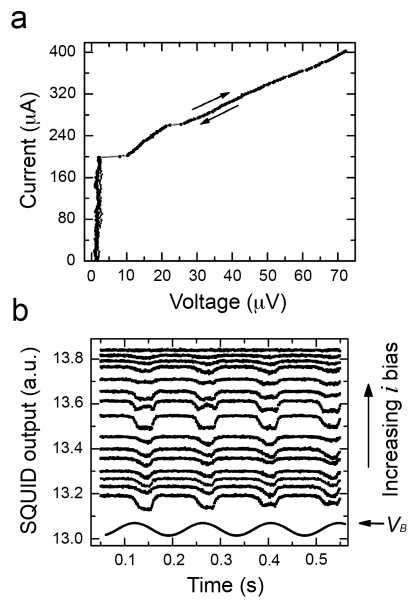

**Figure 2:** (a) *I-V* curves obtained for device A at 6 K. (b) The output voltage of device A, measured at many different values of current bias, in response to a sinusoidal bias, $V_B$, applied to a magnetic field coil.

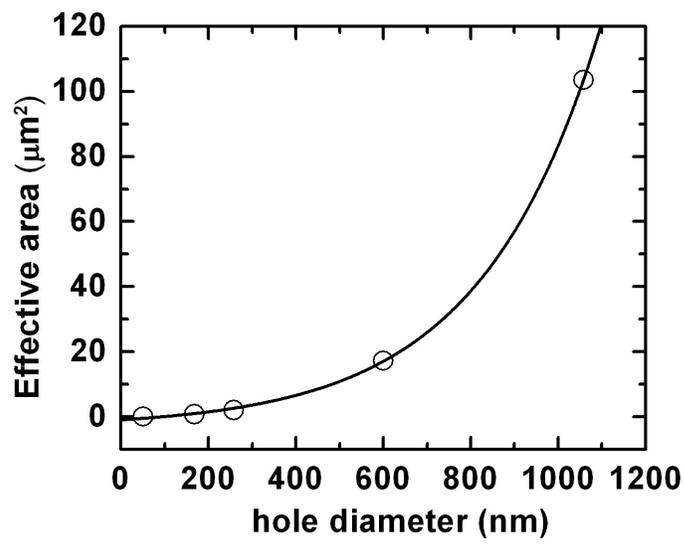

**Figure 3:** A plot of the devices effective area against the hole diameter. The line is a guide of eye.

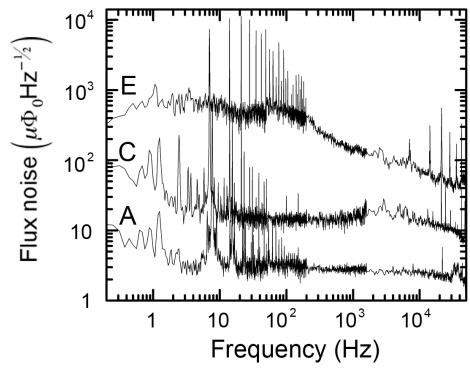

**Figure 4:** (a) Noise spectra of devices A (measured at 6 K), C (4.75K), E (5.3 K). The peak at 8 Hz and its harmonics are the AC field for the transfer function measurement.

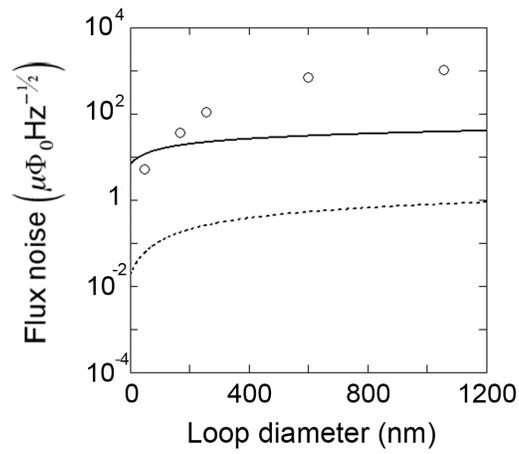

**Figure 5:** Comparison of the measured flux noise at 1 Hz (symbols) with theory based upon electron hopping on defect sites (curves). Solid curve: noise due to defects within the SQUID loop with an areal density of $3 \times 10^{20}$ m$^{-2}$. Broken curve: noise due to defects in the washer with an areal density of $5 \times 10^{17}$ m$^{-2}$.

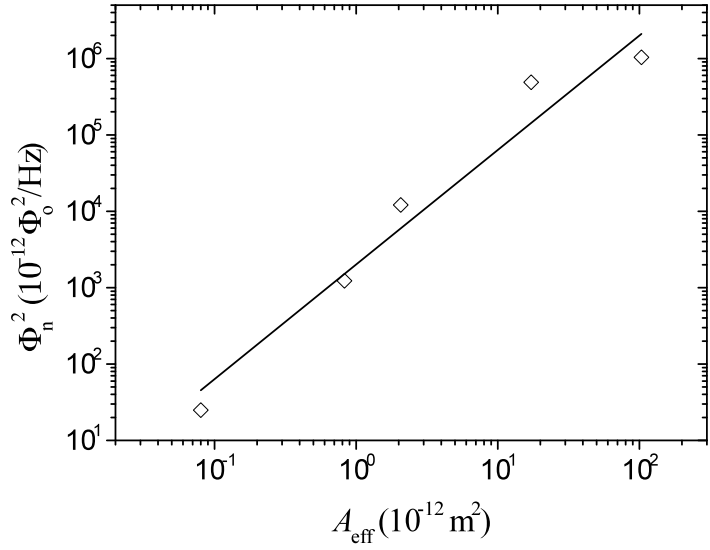

**Figure 6:** A plot of the spectral noise response at 1 Hz against the effective area of the devices (Table 1). The solid line is a fit of the data using a power law $\Phi_n^2 \propto A_{eff}^{1.5}$.